\documentclass[12pt]{iopart}
\usepackage{color}
\usepackage{graphicx}
\usepackage{amsmath, amsthm, amssymb}
\usepackage{enumerate}
\usepackage[normalem]{ulem}
\definecolor{Gray}{gray}{0.7}
\usepackage{citesort}
\usepackage{tabularx}
\usepackage{colortbl}
\usepackage[up]{subfigure}

\newcommand{\be}{\begin{equation}}
\newcommand{\ee}{\end{equation}}
\newcommand{\bea}{\begin{eqnarray}}
\newcommand{\eea}{\end{eqnarray}}

\begin{document}

\title[Shock propagation in hard sphere gas]{Shock propagation in the hard sphere gas in two dimensions:  comparison between simulations and hydrodynamics} 
\author{Jilmy P. Joy$^{1,2,3}$ and R. Rajesh$^{1,2}$}
\ead{jilmyp@imsc.res.in, rrajesh@imsc.res.in}
 \address{The Institute of Mathematical Sciences, C.I.T. Campus,
 Taramani, Chennai 600113, India$^{1}$}
 \address{Homi Bhabha National Institute, Training School Complex, Anushakti Nagar, Mumbai 400094, India$^{2}$}
 \address{Department of Physics, St. Thomas College (Autonomous), Thrissur, Kerala 680001, India$^3$}

\date{\today}

\begin{abstract}
We study the radial distribution of  pressure, density, temperature and flow velocity fields at different times in a two dimensional hard sphere gas that is initially at rest and disturbed by injecting kinetic energy in a localized region through large scale event driven molecular dynamics simulations. For large times, the growth of these distributions are scale invariant. The hydrodynamic description of the problem, obtained from the continuity equations for the three conserved quantities -- mass, momentum, and energy -- is identical to those used to describe the hydrodynamic regime of a blast wave  propagating through a medium at rest, following an intense explosion, a classic problem in gas dynamics. Earlier work showed that the results from simulations matched well with the predictions from hydrodynamics in two dimensions, but did not match well in three dimensions. To resolve this contradiction, we perform large scale simulations in two dimensions, and show that like in three dimensions, hydrodynamics does not describe the simulation data well. To account for this discrepancy, we check in our simulations the different assumptions of the hydrodynamic approach like local equilibrium, existence of an equation of state, neglect of heat conduction and viscosity.
\end{abstract}

\maketitle

\section{\label{introduction} Introduction}

The study of the propagation of a blast wave in a gas caused by the input of a large amount of energy in a localised region of space, is one of the classic problems in gas dynamics~\cite{Whithambook,barenblatbook}. Initially, energy is transported from the location of input to the outside primarily in the form of radiation. As the gas cools with time, radiation becomes less important, and the transport of energy is dominated by a shock wave, in which the perturbed matter moves faster than the speed of sound. In this regime, the expansion becomes self similar in time. The radius of the shock front, from dimensional arguments, increases as $R(t) \sim t^{2/d+2}$ where $d$ is the spatial dimension~\cite{gtaylor1950,gtaylor1950_2,jvneumann1963cw,lsedov_book,sedov1946}. The spatial and temporal dependence of the different thermodynamic quantities like density, pressure, temperature and flow velocity can be obtained using hydrodynamics by studying the continuity equations for conservation of density, momentum and energy. The exact solution for these in three dimensions, when heat conduction and viscous effects are ignored,  were found by Taylor, von Neumann and Sedov~\cite{gtaylor1950,gtaylor1950_2,jvneumann1963cw,lsedov_book,sedov1946},  and we will refer to this theory as the TvNS theory.

Examples of  physical systems where the hydrodynamic regime of a blast wave has been studied include the Trinity nuclear explosion of 1945~\cite{gtaylor1950,gtaylor1950_2},   intermediate time evolution of supernova remnants~\cite{lsedov_book,lwoltjer1972,sfgull1973,dfcioffi1988,jpostriker1988rmp,ybzeldovich_book}, and laser-driven blast waves in gas jets~\cite{mjedwards2001prl}, plasma~\cite{adedensphysplasma2004}, or in cluster media~\cite{asmoorephysplasma2005}. These studies have focused on verifying the scaling law for the growth of the shock front.  Further generalisations and applications of the TvNS theory include the case when there is a continuous input of energy in a localised region~\cite{dokuchaev2002aanda,saegfalle1975astronandastrophys}, inclusion of the effects of heat conduction~\cite{afghoniem1982jfm,amraouf1991fluiddynres,hsteiner1994physfluids} and viscous effects~\cite{neumann1950japplphys,rlatter1955japplphys,hlbrode1955japplphys,mnplooster1970physfluids}. The TvNS theory has also been generalised to examples where the number of conserved quantities are fewer in number. For example, in granular systems, energy is no longer conserved in collisions, while momentum and mass are. There are many situations when the response of a dilute granular system to  localized perturbations, either as an impact or continuous in time, is of interest. Examples  include crater formation in a granular bed following an impact of an object or a continuous jet~\cite{amwalsh2003prl,ptmetzger2009aip,ygrasselli2001granmatt}, shock propagation in a granular medium following a sudden impact~\cite{jfboudet2009prl,zjabeen2010epl,snpathak2012pre}, viscous fingering by the continuous injection of energy~\cite{xcheng2008natphy,bsandnes2007prl,sfpinto2007prl,ojohnsen2006pre,hhuang2012prl}, shock propagation in continuously driven granular media~\cite{jpjoy2017pre}, etc.  More recently, the TvNS theory has 
been generalized to include dissipative interactions in order  to describe the spatial variation of density, temperature, etc., in these systems~\cite{mbarbier2015prl,mbarbier2016physfluids}.

While the hydrodynamic equations and their modifications have been studied in great detail, it has only been more  recently that the theory been tested in simulations of particle based models. The simplest model is the hard sphere gas in which particles move ballistically until they undergo momentum and energy conserving collisions.  If  the particles are initially at rest and energy is imparted to  a few particles in a localised region, then a shock wave is set up. After initial transients, a self similar regime is reached. Since the only conserved quantities in this model are density, momentum, and energy, as in the TvNS theory, it is a direct realisation of the TvNS theory. The power law growth for the shock front has been verified in simulations of such systems both in two~\cite{tantal2008pre,zjabeen2010epl} and three dimensions~\cite{zjabeen2010epl}. For the spatial variation of density, pressure, temperature and flow velocity, the results are not so clear. In two dimensions, for low to medium densities, it was found that simulations reproduce well the TvNS solution for the radial variation of density, flow velocity and temperature fields, except for a small difference in the discontinuities at the shock front, and a slight discrepancy near the shock center~\cite{mbarbier2016physfluids}. However, in three dimensions, from large scale simulations, we showed recently~\cite{jpjoy2018arxiv}  that  the TvNS theory fails to describe the simulation data at most spatial locations, ranging from the shock center to the shock front. In addition, we tested several assumptions of the TvNS theory  within the simulations. It was shown that a key assumption of an existence of an equation of state (EOS) relating the local pressure to the local density and temperature, holds good in simulations. However, while thermal energy is equipartitioned in the different directions, local equilibrium fails to hold. 

Thus, there is a clear discrepancy between the conclusions drawn from the simulations performed in two and three dimensions. Is this because dimension plays a role in the validity of the TvNS theory? Are the assumptions of TvNS theory like local equilibrium, that is invalid in three dimensions, valid in two dimensions? The aim of the paper is to answer these two questions.

In this paper, we perform large scale event driven simulations of shock propagation in the hard sphere gas in two dimensions to test the predictions of the TvNS theory by repeating the analysis that we did for the three dimensional case. In contradiction to the earlier results for two dimensions, we show unambiguously that the simulation data for distances ranging from the shock center to the shock front do not agree with the predictions of the TvNS theory for the hard sphere gas. We also test the key assumptions of the TvNS theory. Like in three dimensions, we find that, there is an EOS relating pressure to density and temperature. This EOS state is the same as that for the hard sphere gas in equilibrium at the local pressure and temperature.  We also find that, as expected for a system in local equilibrium,  energy is equipartitioned equally among the different translational degrees of freedom.   However,  the  distribution of the velocity fluctuations, in regions between the shock center and shock front, is found to have non-gaussian tails. In particular, it is asymmetric with non-zero skewness. These features are also similar as to what was observed in three dimensions.

The remainder of the paper is organized as follows. In Sec.~\ref{sec:model}, we  describe the  model and give details of the simulations. Section~\ref{analytical} describes the hydrodynamic theory for the shock propagating in a two dimensional hard sphere gas, obtained by modifying the TvNS theory to account for steric effects. In Sec.~\ref{comparison} we compare the TvNS predictions for the radial variation of the density, velocity, pressure and temperature with the results obtained from large-scale simulations of the hard sphere gas.  We test the different assumptions of the  TvNS theory within simulations in Sec.~\ref{local-equilibrium}.  Section~\ref{conclusion} contains a summary and discussions.

\section{\label{sec:model} Model}

Consider a collection of stationary hard spheres that are initially uniformly distributed in space. The mass and diameter are set to one. The system is perturbed by an isotropic impulse at the origin. To model an isotropic impulse in the simulations, we choose four particles near the origin and assign  to them velocities of magnitude $1$ along the $\pm x$ and $\pm y$ directions. The particles move ballistically and transfer kinetic energy to other particles through energy and momentum conserving collisions.
If $\vec{u}{_1}$ and $\vec{u}{_2}$ are the pre-collision velocities of two particles $1$ and $2$, then the corresponding post-collision velocities 
$\vec{v}{_1}$ and $\vec{v}{_2}$, are given by
\bea
\vec{v}{_1} &=& \vec{u}{_1} -  [\hat{n} \cdot (\vec{u}{_1}-\vec{u}{_2})] \hat{n}, \nonumber \\
\vec{v}{_2} &=& \vec{u}{_2} -  [\hat{n} \cdot (\vec{u}{_2}-\vec{u}{_1})] \hat{n},
\eea
where $\hat{n}$ is the unit vector along the line joining the centers of particles $1$ and $2$. The only control parameter in the problem is the initial number density $\rho_0$.

\begin{figure}
\centering
\includegraphics[width=0.8\columnwidth]{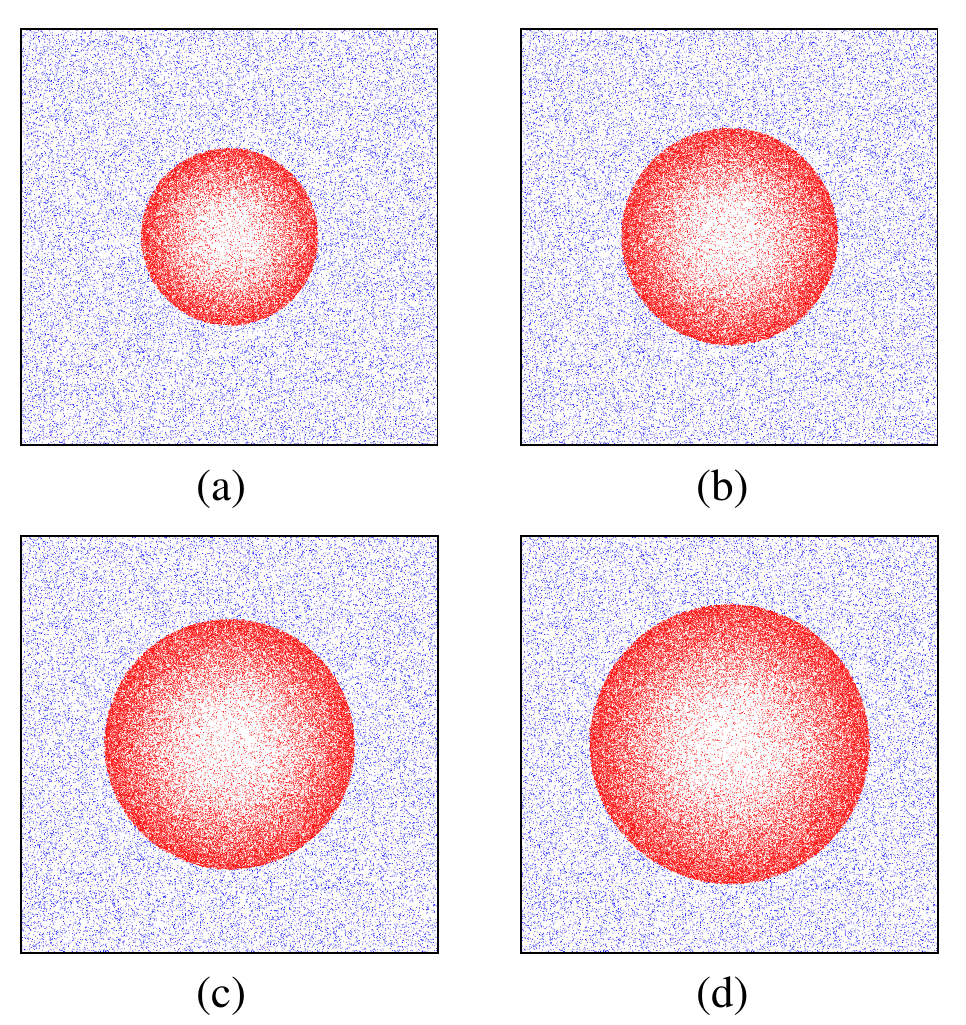}
\caption{ Moving (red) and stationary (blue) particles at times (a) $t = 1\times 10^5$, (b) $t = 1.5 \times 10^5$, (c) $t = 2.0 \times 10^5$ and  (d) $t = 2.5 \times 10^5$, after the initial injection of  four energetic particles  at the center. At time $t=2.5 \times 10^5$, there are $919407$ moving particles. The data are for the ambient number density $\rho_0=0.15$.}
\label{fig:snapshot-single-impact-2d}
\end{figure}

We simulate the system using event driven molecular dynamics simulations~\cite{dcrapaportbook}. 
The initial perturbation creates a disturbance, made up of moving particles,  that propagates radially outwards. Snapshots of the system at different times are shown in Fig.~\ref{fig:snapshot-single-impact-2d}. The moving particles are separated from the stationary particles by a shock front. 

\begin{figure}
\centering
\includegraphics[width=\columnwidth]{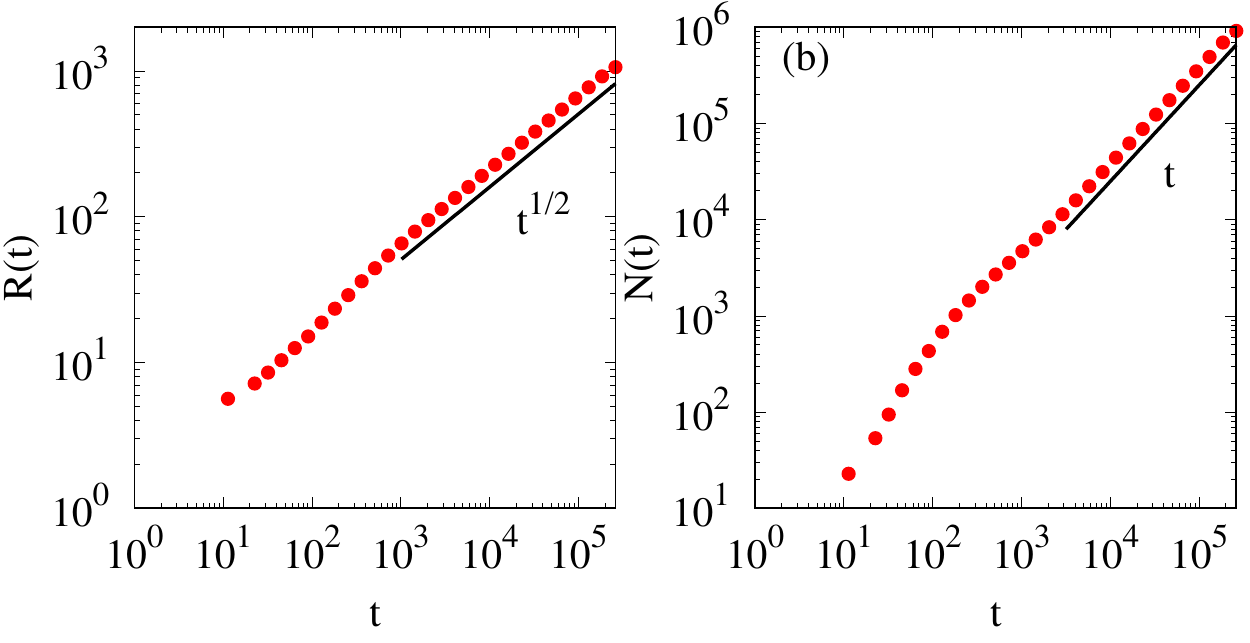}
\caption{Simulation results for the temporal variation of (a) the radius of the shock $R(t)$ and  (b) the number of moving particles $N(t)$. The solid lines are power laws (a) $\sqrt{t}$ and (b) $t$. The data are for the ambient number density $\rho_0=0.15$.}
\label{fig:power-law-benchmark}
\end{figure}

The radius of this shock front  $R(t)$ has been  shown earlier to increase as $t^{1/2}$ in event driven simulations, consistent with dimensional analysis~\cite{tantal2008pre,zjabeen2010epl}. To benchmark our simulations as well as to estimate the time for initial transients, we show in Fig.~\ref{fig:power-law-benchmark} the temporal variation of both $R(t)$ and  the total number of moving particles $N(t)$.  We define $R(t)$ as the radius of gyration of the moving particles at a given time $t$.  In Fig.~\ref{fig:power-law-benchmark}, at large times, we find $R(t) \sim \sqrt{t}$, and $N(t)$ increases as $t$, consistent with $N(t) \sim R(t)^2$. However, it can be seen that there are strong initial transients before the asymptotic behaviour is attained. For studying the scaling behaviour of the different thermodynamic quantities, we choose times that are larger than this crossover time. 

In our simulations, we measure the radial variation of pressure, density, temperature and flow velocity by averaging the simulation data 
over $150$ different histories.
We simulate systems with two different number densities $\rho_0 = 0.15, 0.382$, both of which are much smaller than the random closed packing density.  To ensure that there are no boundary effects, the number of particles and time of simulation are chosen such that the moving particles are far from the  boundary.  The local temperature is measured from the velocity fluctuations, obtained by subtracting out the mean radial velocity from the instantaneous velocity. The local pressure is measured from the local collision rate. For the hard sphere gas in two dimensions,  pressure is given by~\cite{masaharumdsimulation2016} 
\be
 p= \rho T - \frac{\rho}{2 N \delta t} \sum_{collisions}b_{ij},
 \label{eqn:pressure}
\ee
where $b_{ij}=\vec{r}_{ij} \cdot \vec{v}_{ij}$, where $\vec{r}_{ij}$ and $\vec{v}_{ij}$ respectively are the relative positions and velocities of the particles $i$ and $j$ undergoing collisions, $\delta t$ is the time duration of measurement, and $N$ is the mean number of particles in the radial bin where pressure is being computed.

\section{\label{analytical} Hydrodynamics}

In this section, we describe the TvNS theory for the hydrodynamical description of shock propagation following an intense, isotropic, localized perutrbation, modified to include steric effects due to the finite sizes of the spheres. Initially, the gas that is at rest with  number density $\rho_0$, is perturbed by adding energy $E_0$ at the center.  The mass, momentum, and energy are conserved locally so that the fluid flow is described by the corresponding continuity equations. In the TvNS theory, it is assumed that heat conduction and viscous effects may be ignored and that local equilibrium is achieved.  These assumptions imply  that the flow is isentropic. Thus, the conservation law for energy can be replaced by that for entropy. Since the flow is isotropic, the different thermodynamic quantities cannot depend on the angle. Thus, in radial coordinates, the continuity equations are~\cite{Landaubook}
\begin{align}
 &\partial_t \rho+\partial_r(\rho v)+  r^{-1}\rho v=0, \label{eq:conservation-density-2d} \\ 
 &\partial_t v+v \partial_r v+ \rho^{-1}\partial_r p=0,  \label{eq:conservation-momentum-2d}\\ 
 &\partial_t s+v \partial_r s=0,  \label{eq:conservation-entropy-2d}
\end{align}
where $\rho$ is the density, $v$ is the mean radial velocity, $p$ is the pressure and $s$ is the entropy.

The number of independent parameters are reduced by assuming local equilibrium. This  implies that the local pressure is related to the local density and temperature through an  EOS. Here, temperature is a measure of the local velocity fluctuations about the mean flow velocity. In the original TvNS theory, the EOS was chosen to be  that of the ideal gas, making the resulting equations solvable. For the hard sphere gas, steric effects are important. To include these effects, more realistic virial EOS was used in three dimensions~\cite{jpjoy2018arxiv}, and the Henderson EOS was used in two dimensions~\cite{mbarbier2016physfluids}. We now describe the hydrodynamics with virial EOS in two dimensions, and discuss the role of truncation of  the virial expansion.

The EOS of a  gas has the virial expansion
\begin{align}
&\frac{p}{k_B T \rho}= 1+ \sum_{n=2}^{\infty} B_n \rho^{n-1}, \label{eqn:virial} 
\end{align}
where $T$ is the  temperature, $k_B$ is the Boltzmann constant, and $B_n$ are the virial coefficients. The entropy as a virial expansion is then  given by
\be
 s =N k_B \left[\frac{3}{2} - \ln(\Lambda^2 \rho) - \sum_{n=2}^{\infty} \frac{\rho ^{n-1}}{n-1} \left( B_n + T \frac{d B_n}{d T}\right)\right], \label{eqn:entropy_virial}
\ee 
where $\Lambda = h/\sqrt{2 \pi m k_B T}$ is the thermal wavelength. For hard spheres, the virial coefficients are independent of temperature, i.e.,  $dB_n/dt=0$. The virial coefficients $B_n$ for the hard sphere gas in two dimensions  are known analytically for up to $n=4$ and through Monte Carlo simulations up to $n=10$~\cite{McCoybook}. These are tabulated in  Table~\ref{table:Bn}.
\begin{table}
\caption{\label{table:Bn} 
The values of the virial coefficients $B_n$ for the hard sphere gas in two dimensions. The data are taken from Ref.~\cite{McCoybook}. } 
\begin{indented}
\lineup
\item[]\begin{tabular}{@{}ll}
\br
$n$  & $B_n$  \\
\mr
 $2$ &  $\frac{\pi}{2}$\\
 $3$ & $ (\frac{4}{3}-\frac{\sqrt{3}}{\pi}) B_2^2$\\
 $4$ & $\Big[2-\frac{9 \sqrt{3}}{2 \pi}+\frac{10}{\pi ^2}\Big] B_2^3$\\ 
 $5$ & $0.33355604 B_2^4$\\
 $6$ & $0.1988425 B_2^5$\\
 $7$ & $0.11486728 B_2^6$\\
 $8$ & $0.0649930 B_2^7$\\
 $9$ & $0.0362193 B_2^8$\\
 $10$ & $0.0199537 B_2^9$\\
   \br
  \end{tabular}
  \end{indented}
\end{table}

Substituting the virial expansions for  pressure and entropy in Eqs.~(\ref{eq:conservation-density-2d})-(\ref{eq:conservation-entropy-2d}), we obtain
\begin{align}
&\partial_t\rho+\partial_r(\rho v)+  r^{-1}\rho v=0, \label{eqn:virial_conservation_laws-mass-2d}\\ 
&(\partial_t + v \partial_r)v + k_B T \left[ 1 + \sum_{n=2}^{\infty} n B_n \rho^{n-1} \right] \partial_r \ln \rho + k_B T \left[ 1 + \sum_{n=2}^{\infty} B_n \rho^{n-1} \right] \partial_r \ln T = 0, \label{eqn:virial_conservation_laws-momentum-2d}\\
&(\partial_t + v \partial_r)   \ln T-\left[ 1\! +\! \sum_{n=2}^{\infty} B_n \rho^{n-1} \right]\!(\partial_t + v \partial_r)  \ln \rho = 0.
  \label{eqn:virial_conservation_laws-entropy-2d}
\end{align}
Non-dimensionalising the different thermodynamic quantities converts  Eqs.~(\ref{eqn:virial_conservation_laws-mass-2d})--(\ref{eqn:virial_conservation_laws-entropy-2d}) from partial to  ordinary differential equations. From dimensional analysis~\cite{barenblatbook}
\bea
p&=&\frac{\rho_0 r^2}{t^2}P(\xi), \nonumber \\
\rho&=&\rho_0 R(\xi), \label{eqn:dimensional_analysis-2d} \\
v&=&\frac{r}{t}V(\xi), \nonumber\\
\varepsilon&=& \frac{k_B T}{m_0} = \frac{r^2}{t^2}E(\xi) \nonumber,
\eea
where
\be
 \xi=r\left(\frac{E_0 t^2}{\rho_0}\right)^{-1/4},
 \label{eqn:independent_variable-2d}
\ee
is the non-dimensionalised length, $E_0$ is the initial energy that is input at the spatial location $r=0$, $\rho_0$ is the ambient mass density, $T$ is the local temperature, $k_B$ is Boltzmann constant, $m_0$ is the mass of a particle, and $P$, $R$, $V$, and $E$, are scaling functions. $\varepsilon$ is the thermal energy per unit mass. The four scaling functions are related through the virial EOS [see Eq.~(\ref{eqn:virial})] as
\be
P(\xi)=E(\xi) R(\xi)\left[1 + \sum_{n=2}^{\infty} B_n \rho_0^{n-1} R(\xi)^{n-1}\right].
 \label{eqn:virial_equation_of_state}
\ee
Equations~(\ref{eqn:virial_conservation_laws-mass-2d})--(\ref{eqn:virial_conservation_laws-entropy-2d}) may be rewritten in terms of the scaling functions as  
\begin{align}
\left(V-\frac{1}{2}\right) R \xi  \frac{dV}{d\xi} + \xi  \frac{d}{d\xi} \left[ER \left(1+ \sum_{n=2}^{\infty} B_n \rho_0^{n-1} R^{n-1} \right)\right]
 - R V + R V^2 \nonumber \\+ 2 R E\left[1 + \sum_{n=2}^{\infty} B_n \rho_0^{n-1} R^{n-1}\right]=0, \label{eq:ode1}\\
\left(V-\frac{1}{2}\right) \xi \frac{dR}{d\xi} + \xi R \frac{dV}{d\xi} + 2 R V=0,\\
 -\Big(1 + \sum_{n=2}^{\infty} B_n \rho_0^{n-1} R^{n-1}\Big) \left(V-\frac{1}{2}\right) \frac{\xi}{R} \frac{dR}{d\xi} +  \left(V-\frac{1}{2}\right) \frac{\xi}{E} \frac{dE}{d\xi} +2 (V-1)=0.
  \label{eqn:ode3}
\end{align}

The various thermodynamic quantities are discontinuous across the shock front. These discontinuities are determined based on the flow of conserved quantities across the shock front and given by the Rankine-Hugoniot boundary conditions~\cite{Landaubook}.
The Rankine-Hugoniot boundary conditions at the shock front $\xi_f$ in terms of dimensionless variables are
\begin{align}
&\frac{1}{R(\xi_f )}\left[1+\frac{2}{1+ \sum_{n=2}^{\infty} B_n \rho_0^{n-1} R(\xi_f )^{n-1}}\right]=1,\nonumber \\
&V(\xi_f )=  \frac{1}{R(\xi_f )[1+ \sum_{n=2}^{\infty} B_n \rho_0^{n-1} R(\xi_f)^{n-1}]}, \nonumber\\
&E(\xi_f ) =\frac{1}{2} V(\xi_f )^2. \label{eq:rankine-virial}
\end{align}

For a given $\xi_f$, Eqs.~(\ref{eq:ode1})-(\ref{eqn:ode3}) with the boundary conditions in Eqs.~(\ref{eq:rankine-virial}) may be solved numerically. $\xi_f$  is then determined by the condition that total energy is conserved. This constraint, in  terms of the scaling functions, is
\be
2 \pi \int_0^{\xi_f}R(\xi)\left[\frac{V^2(\xi)}{2}+ E(\xi) \right]\xi^3 d\xi = 1.
\label{eqn:xif_constraint-virial}
\ee
To obtain the numerical solution to the set of ODEs [see Eqs.~(\ref{eq:ode1})--(\ref{eqn:ode3})], we convert  this boundary value problem to an initial value problem by choosing a numerical value of $\xi_f$.   The value of $\xi_f$ is iterated till the solution satisfies Eq.~(\ref{eqn:xif_constraint-virial}) within a pre-determined accuracy.  

\begin{figure}
 \centering
 \includegraphics[width=0.4\columnwidth]{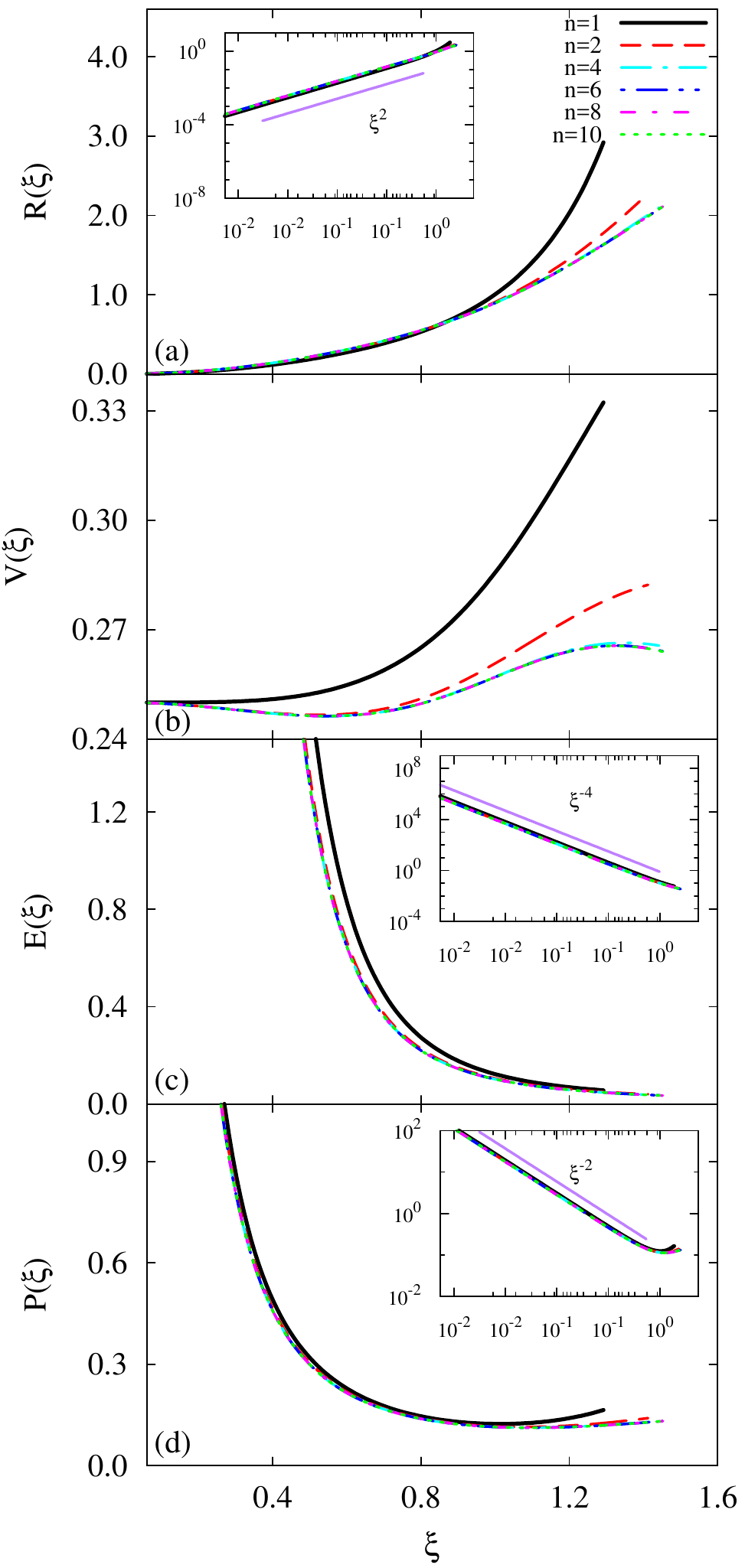}
 \includegraphics[width=0.4\columnwidth]{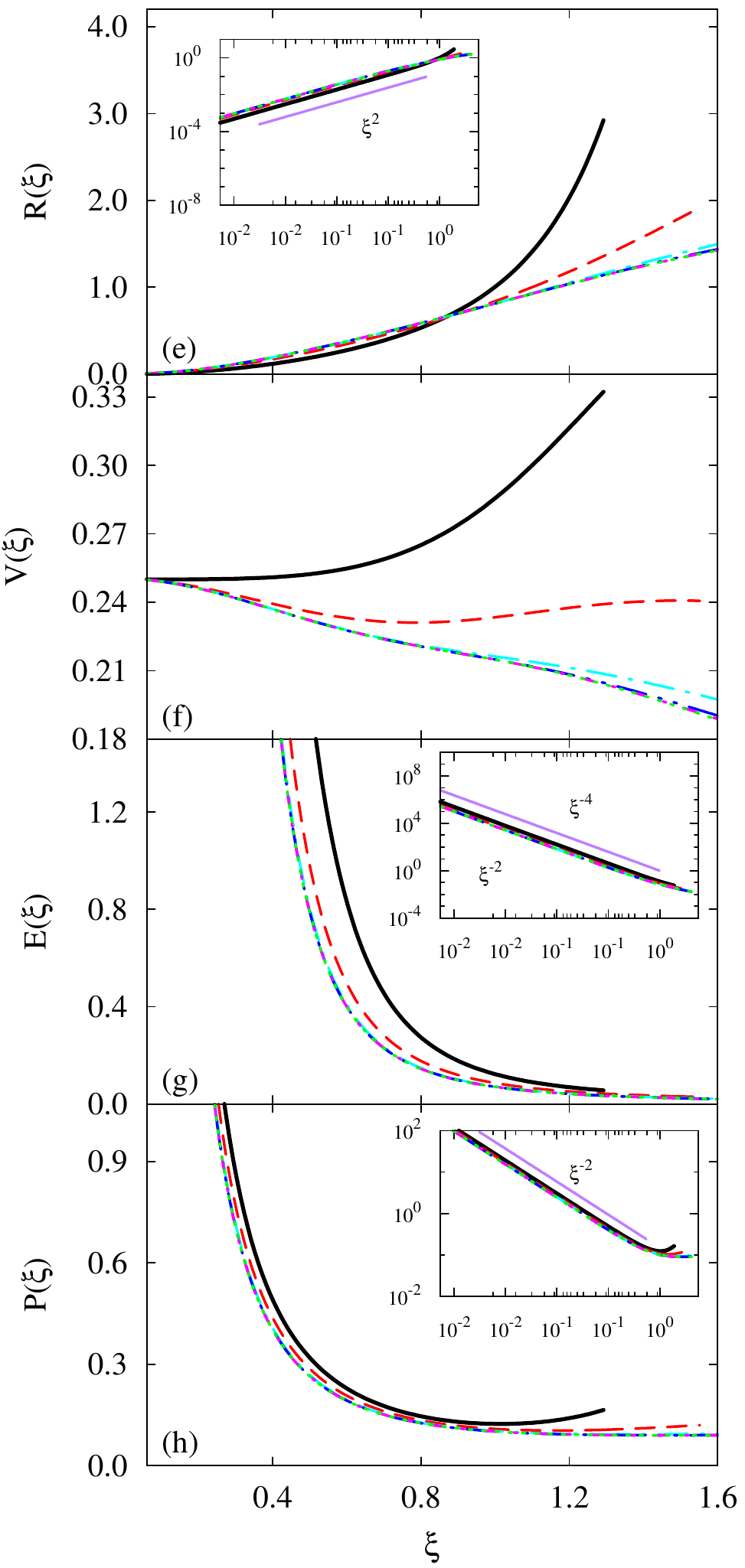}
 \caption{(Color online) The scaling functions (a) $R(\xi)$, (b) $V(\xi)$, (c) $E(\xi)$, and (d) $P(\xi)$ corresponding to density, velocity, temperature and pressure respectively versus $\xi$ obtained from hydrodynamic equations for ambient number density (a)--(d) $\rho_0=0.15$ and (e)--(h) $\rho_0=0.382$.  $n$ refers to the number of terms that is retained in the virial expansion ($n=1$ is ideal gas).  The insets show the plots on a log-log scale, accentuating the small $\xi$ behavior. } 
 \label{fig:scaling_functions_virial}
\end{figure}

The numerically obtained scaling functions are  shown in Fig.~\ref{fig:scaling_functions_virial} for ambient number densities $\rho_0=0.15$ and $\rho_0=0.382$. The different curves correspond to the number of terms that are retained  in the virial EOS ($n=1$ corresponds to the ideal gas EOS).  Three features may be deduced from the data. First is that the ambient number density $\rho_0$ affects the scaling functions. Second is that the data for $n=6$ can hardly be distinguished from that for $n=10$ for both ambient number densities. This means that, though the virial coefficients are known only upto $n=10$, they provide a very good approximation to the actual EOS, for the number densities that we will be working with. Third, the exponents characterising the small $\xi$ power law behaviour of $R$, $E$, and $P$ are robust and independent of the EOS.

\section{\label{comparison} Comparison of hydrodynamics with simulations}

\begin{figure}
\centering
 \includegraphics[width=0.85\columnwidth]{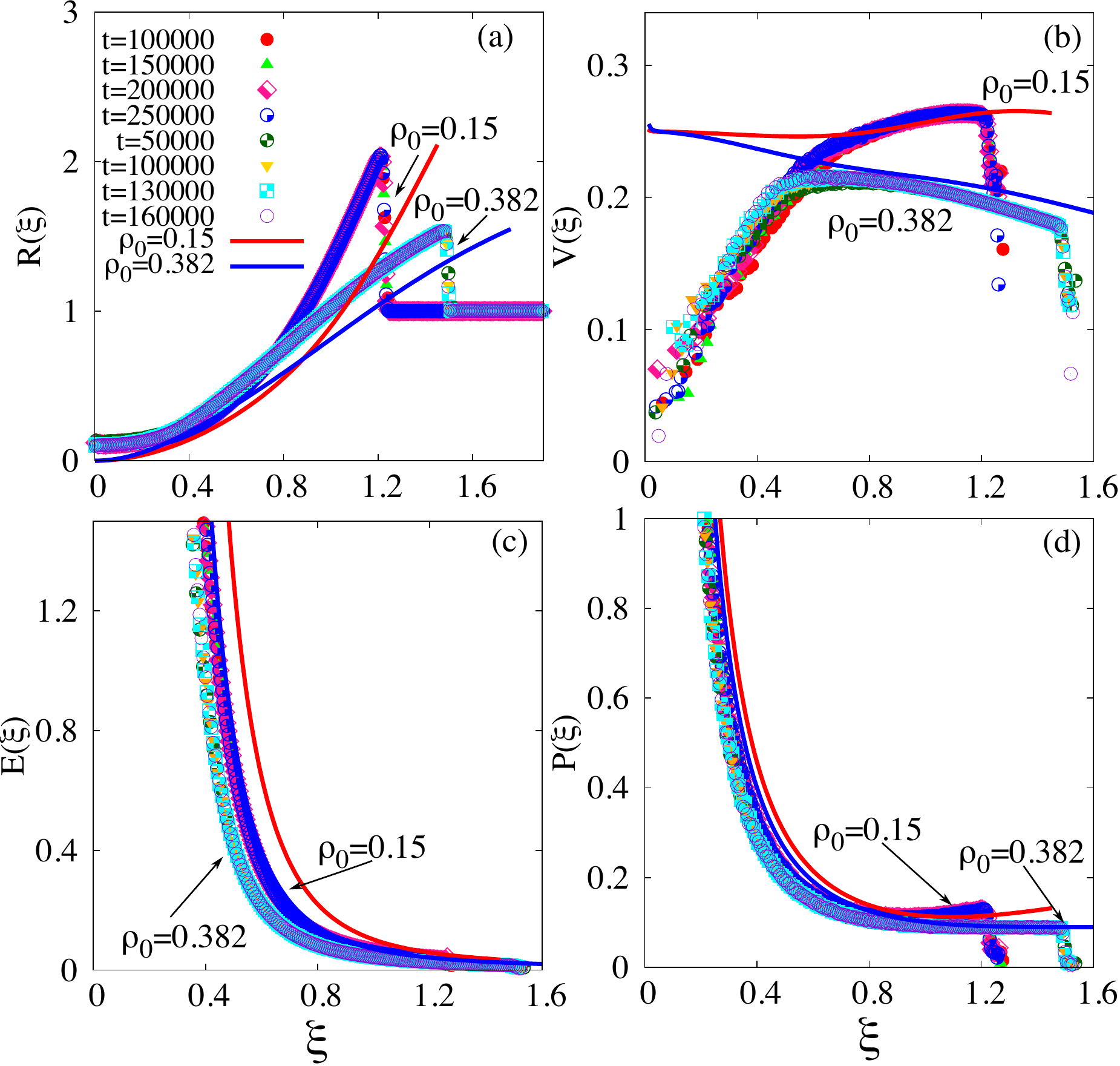}
 \caption{(Color online)  The variation of the scaling functions (a) $R(\xi)$, (b) $V(\xi)$, (c) $E(\xi)$  and (d) $P(\xi)$ corresponding to non-dimensionalised density, velocity, temperature and pressure  [see Eq.~(\ref{eqn:dimensional_analysis-2d})] with scaled distance  $\xi$. The data are shown for $2$ different initial densities $\rho_0 =0.15$ and $0.382$. For $\rho_0=0.15$, the different times are $t=100000$, $150000$, $200000$, $250000$, and for $\rho=0.382$, $t=50000$, $100000$, $130000$, $160000$,  as indicated in (a).  The solid lines  correspond to  predictions from the TvNS theory when the virial EOS is  truncated at $n=10$. The data for $R$, $P$, and $E$ are also shown on a logarithmic scale  in Fig.~\ref{fig:data_analysis_comparison_powerlaw}. }
 \label{fig:data_analysis_comparison}
\end{figure}

The  scaling functions $R(\xi)$, $V(\xi)$, $E(\xi)$, and $P(\xi)$ obtained from event driven simulations are shown in Fig.~\ref{fig:data_analysis_comparison} for initial number  densities $0.15$ and $0.382$. For each of the densities, four different times are shown. The data for the different times collapse onto one curve when plotted against $\xi$. The predictions from TvNS solution, when the virial EOS is truncated at the tenth term, are shown by solid lines. 

\begin{figure}
\centering
 \includegraphics[width=\columnwidth]{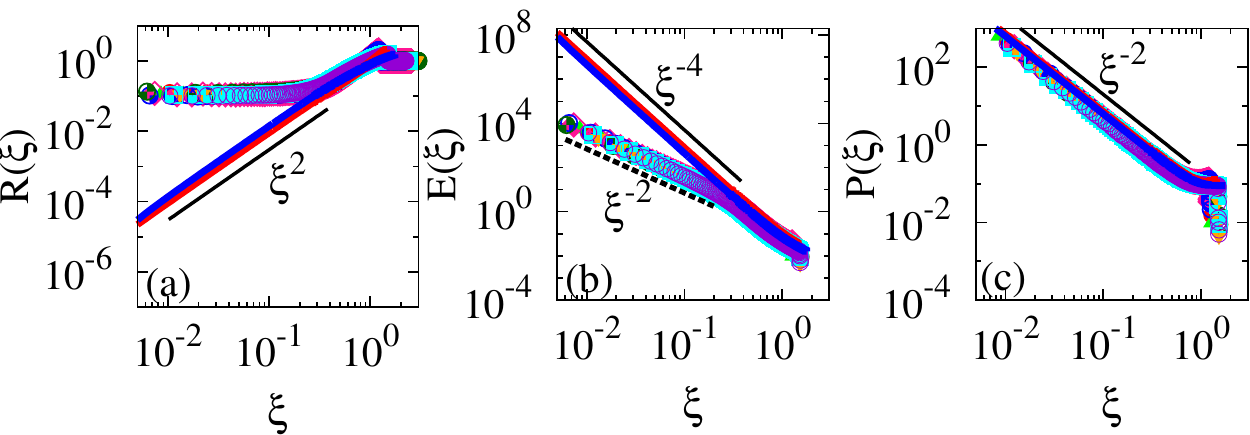}
 \caption{(Color online)  The data in Fig.~\ref{fig:data_analysis_comparison}(a), (c) and (d) are shown in logarithmic scale to emphasize the power-law divergence for small $\xi$. The three panels show the variation of  the scaling functions (a) $R(\xi)$, (b) $E(\xi)$  and (c) $P(\xi)$ with scaled distance  $\xi$. The data are for $2$ different initial densities $\rho_0 =0.15$ and $0.382$. Each density has data for four different times and the symbols are same as described in Fig.~\ref{fig:data_analysis_comparison} (a). The black solid lines correspond to the TvNS solution. }
 \label{fig:data_analysis_comparison_powerlaw}
\end{figure}

All the scaling functions,  especially close to the shock front, depend  on the  ambient number density $\rho_0$. As $\rho_0$ increases, the discontinuity at the shock front  decreases. Most importantly, the TvNS solution does not describe the simulation data well. For the scaling function 
$R(\xi)$ the theorertical and numerical answers do not match for all values of $\xi$.  In particular, as shown in Fig.~\ref{fig:data_analysis_comparison_powerlaw}(a), the TvNS prediction for  $R(\xi)$  increases as a power law $\xi^{2}$ for small $\xi$ while the numerically obtained scaling $R(\xi)$ tends to a non-zero constant.

The scaling function $V(\xi)$, shown in Fig.~\ref{fig:data_analysis_comparison}(b), increases linearly from zero, reaches a maximum and then decreases to its value at the shock front. The  TvNS solution captures the simulation data close to the shock front. However, for smaller $\xi$, the TvNS solution for $V(\xi)$ tends to a non-zero constant, while the simulation results show that $V(\xi)$ tends to zero for small $\xi$.
The scaling function $E(\xi)$, which measures the square of the local velocity fluctuations, is shown in Fig.~\ref{fig:data_analysis_comparison}(c). There is only a weak  dependence on the ambient number density $\rho_0$. From Fig.~\ref{fig:data_analysis_comparison_powerlaw}(b), it can be seen that $E$ diverges  as a power law $E(\xi)\sim \xi^{-2}$ as $\xi \to 0$. However, the TvNS solution predicts that $E$ diverges  
as $E\sim \xi^{-4}$ as $\xi \to 0$, showing a mismatch. 
The dependence of the scaled pressure on $\xi$ is shown in
Fig.~\ref{fig:data_analysis_comparison}(d). Unlike the other scaling
functions, the TvNS solution is a good characterisation of the simulation data.  In particular, both the theoretical predictions as well as the numerical data  diverge as $\xi^{-2}$ as $\xi \to 0$
[see Fig.~\ref{fig:data_analysis_comparison_powerlaw}(c)]. These results, showing a mismatch between the TvNS solution and the numerical data, is quite similar to what was seen in three dimensions~\cite{jpjoy2018arxiv}.

In summary, the  TvNS solution fails  to describe well the numerical data. There are multiple plausible reasons for the observed differences.   Shock propagation is  inherently a system out of equilibrium, and thus the assumption of local equilibrium may be incorrect. Likewise, viscous effects are ignored. In the following, we test these assumptions.

\section{\label{local-equilibrium} Verifying the assumptions of the TvNS theory}

We now numerically check the different assumptions of the TvNS theory.

\subsection{Equation of state}

One of the key assumptions  of the TvNS theory is an EOS relates  the local pressure to the local  density and temperature. To test the assumption of EOS, we independently measure the local thermodynamic quantities numerically and check whether they obey the  hard sphere virial EOS by numerically measuring the ratio
\be
\chi(n)=   \frac{P(\xi)}{E(\xi) R(\xi)\left[1 + \sum_{k=2}^{n} B_k \rho_0^{k-1} R(\xi)^{k-1}\right]},
 \label{eqn:scaling_fn_virial}
\ee
where $n$ is the number of terms retained in the virial expansion [$n=1$ corresponds to ideal gas]. If $\chi \approx 1$ for increasing $n$, then we conclude that the local thermodynamic quantities obey the virial EOS, and hence the assumption of EOS is justified.

\begin{figure}
 \centering
 \includegraphics[width=0.7\columnwidth]{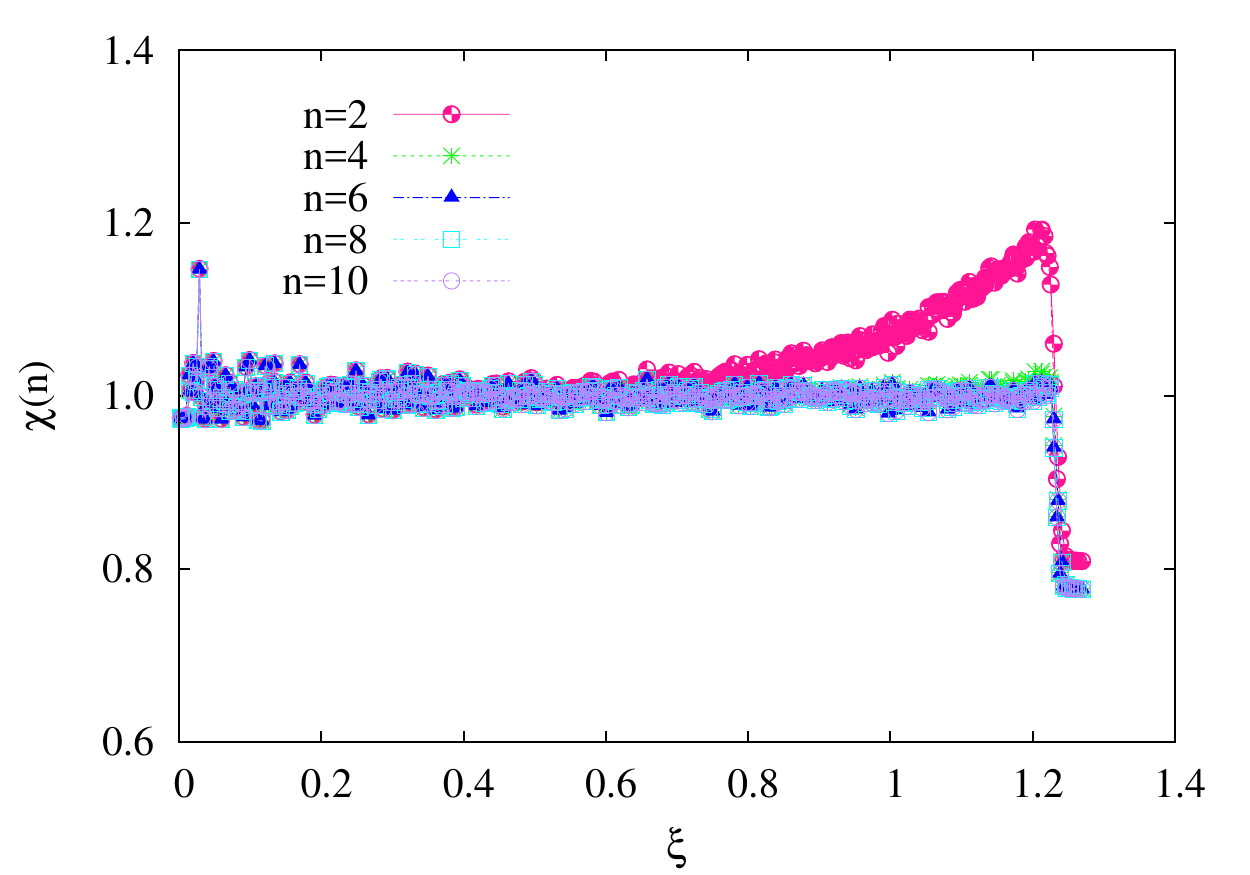}
 \caption{(Color online)  The variation of $\chi(n)$ [see Eq.~(\ref{eqn:scaling_fn_virial})] with $\xi$ for $n=2,4,6,8, 10$. The data are for  times $150000$ and $250000$ and for ambient number density $\rho_0=0.15$.  For large $n$, $\chi(n)$ converges to one. }
 \label{fig:eqn_of_state}
\end{figure}

The dependence of $\chi(n)$ on $\xi$  is shown in Fig.~\ref{fig:eqn_of_state} for $n=2,4,6,8,10$ and for two different times. For small $n$, $\chi(n)$ deviates from one near the shock front. However, quite remarkably, as $n$ increases, $\chi(n)$ converges to $1$ for all $\xi$. We thus conclude that  the assumption of existence of  EOS in the TvNS solution is  justified. 

\subsection{Equipartition}

\begin{figure}
  \centering
 \includegraphics[width=0.8\columnwidth]{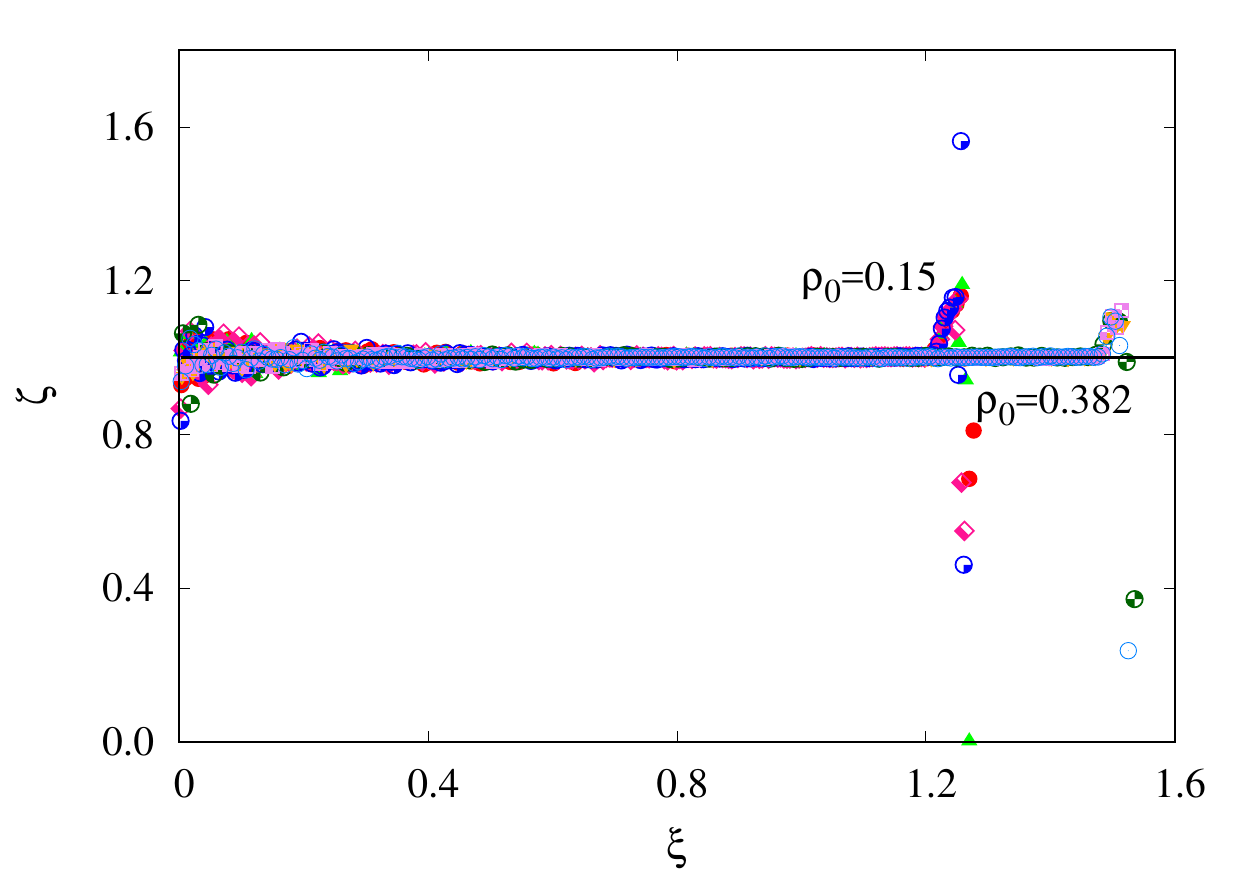}
 \caption{(Color online) The variation of $\zeta$, the ratio of thermal energies in the radial and transverse directions [see Eq.~(\ref{eq:zeta})] with the scaled distance $\xi$. The data is for four different times with keys as  in Fig.~\ref{fig:data_analysis_comparison}(a), for two ambient densities $\rho_0=0.15$ and $0.382$. Away from the shock front, $\zeta \approx 1$.}
 \label{fig:equipartition}
\end{figure}

We check whether the thermal energy is equally equipartitioned into the two degrees of freedom by measuring the ratio
\be
\zeta = \frac{ \langle \delta v_r^2 \rangle}{ \langle \delta v_\perp^2 \rangle},
\label{eq:zeta}
\ee
where $\delta v_r$ and $\delta v_\perp$ are the velocity fluctuations in the radial and transverse directions respectively. When the thermal energy is equipartitioned, then $\zeta$ equals one. The variation of $\zeta$ with $\xi$ is shown in Fig~\ref{fig:equipartition} for different times. The data for different times collapse on to a single curve. We find that $\zeta \approx 1$, except for very close to the shock front, thus showing equipartition. However, near the shock front, $ \zeta >1$, corresponding to excess thermal energy in the radial direction. 

\subsection{Skewness and Kurtosis}

\begin{figure}
 \centering
 \includegraphics[width=0.8\columnwidth]{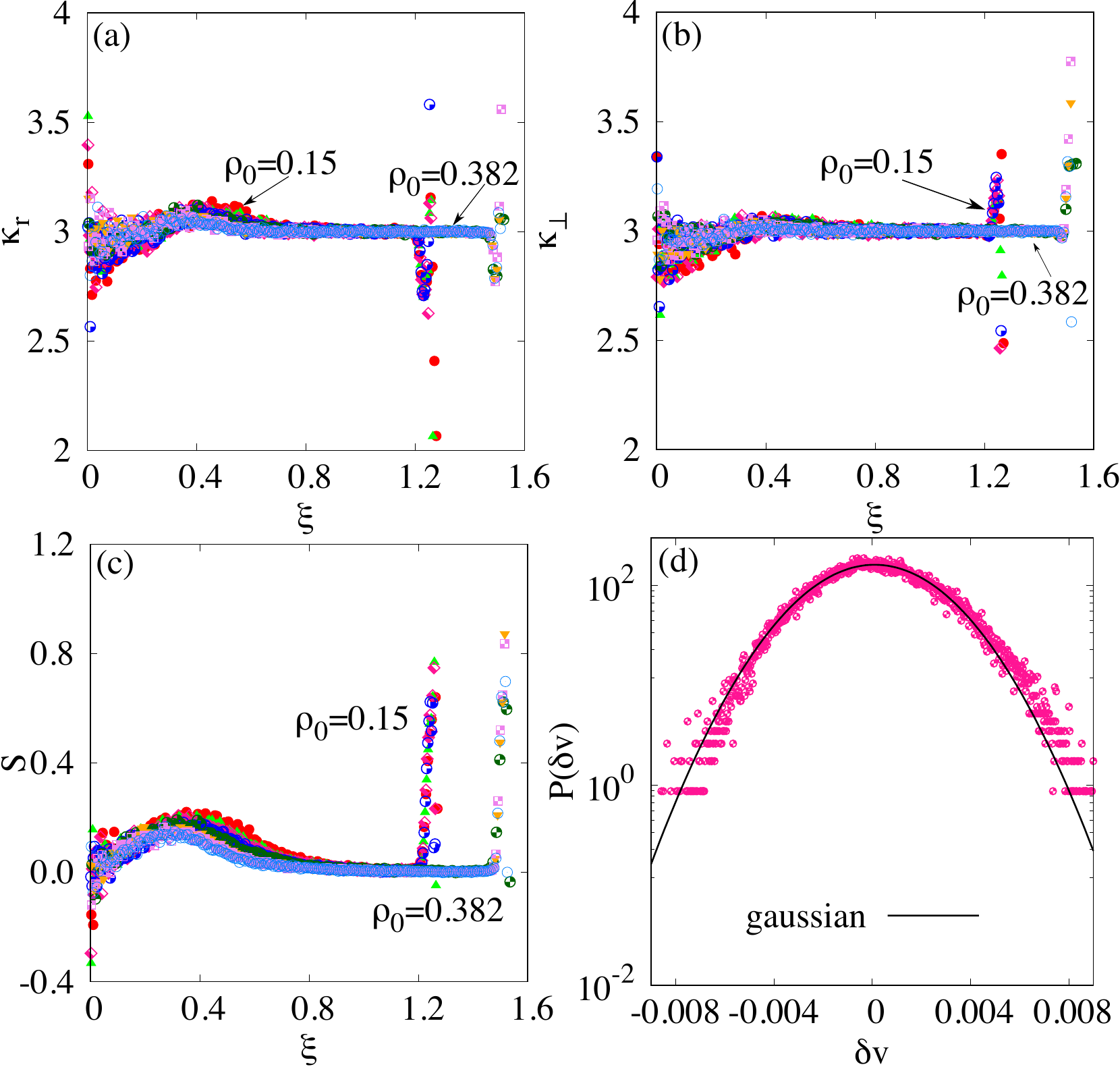}
 \caption{(Color online) The variation with scaled distance $\xi$ of (a) the kurtosis $\kappa_r$ for the radial velocity fluctuations. (b) the  kurtosis $\kappa_\perp$ for the  velocity fluctuations in the $\theta$ direction and (c) skewness $S$ for the radial velocity fluctuations. 
 The data are for  $\rho_0 =0.15$ and $\rho_0=0.382$ and for  four different times  with keys as  in Fig.~\ref{fig:data_analysis_comparison}(a). (d) The distribution of the radial velocity fluctuations $P(\delta v)$ measured at $r=375$, $t=250000$ and $\rho_0=0.15$, corresponding to $\xi=0.33$. The black solid curve represents the best fit of the data to a gaussian distribution.} 
 \label{fig:local_thermal_equilibrium}
\end{figure}

The deviation from gaussianity of the probability distribution for velocity fluctuations can be quantified by measuring the kurtosis $\kappa$ and skewness $S$:
\bea
\kappa_{r, \perp} & = & \frac{\langle \delta v_{r, \perp} ^4\rangle}{ \langle \delta v_{r, \perp} ^2\rangle^2}, \label{eq:kurtosisr-2d}\\
S &=& \frac{\langle \delta v_r^3\rangle}{\langle \delta v_r^2\rangle^{3/2}}. \label{eq:skewness-2d}
\eea
For a gaussian distribution, the kurtosis is $3$, and skewness is zero. Deviation from these values show the non-gaussian behavior. The radial and transverse components of kurtosis are denoted by $\kappa_r$ and $\kappa_\perp$ respectively and their variation with $\xi$ is shown in Fig.~\ref{fig:local_thermal_equilibrium} (a) and (b) respectively. While the data for different times collapse onto one curve, $\kappa_r$  deviates  from $3$ near the shock center, showing a lack of local equilibrium. However,  $\kappa_\perp\approx 3$  for nearly  all $\xi$. In addition to this non-gaussian behaviour, we find that the distribution for the velocity fluctuations is not symmetric with non-zero skewness $S$ for values of $\xi$ close to the shock centre [see Fig.~\ref{fig:local_thermal_equilibrium} (c)].  Thus, the distribution is clearly asymmetric.

To directly observe the skewness of the distribution, we calculate the probability distribution $P(\delta v_r, r, t)$ for the fluctuations of the radial velocity. Figure~\ref{fig:local_thermal_equilibrium}(d) shows the distribution for a fixed time $t$ and  $\xi=0.33$, corresponding to  a region away from the shock front where the skewness in Fig.~\ref{fig:local_thermal_equilibrium}(c) is non-zero. The distribution is compared with the fit to a gaussian. Clearly, the distribution deviates from a gaussian, is asymmetric, and is  skewed towards the larger positive fluctuations.

\subsection{Energy of mean flow}

\begin{figure}
  \centering
 \includegraphics[width=0.8\columnwidth]{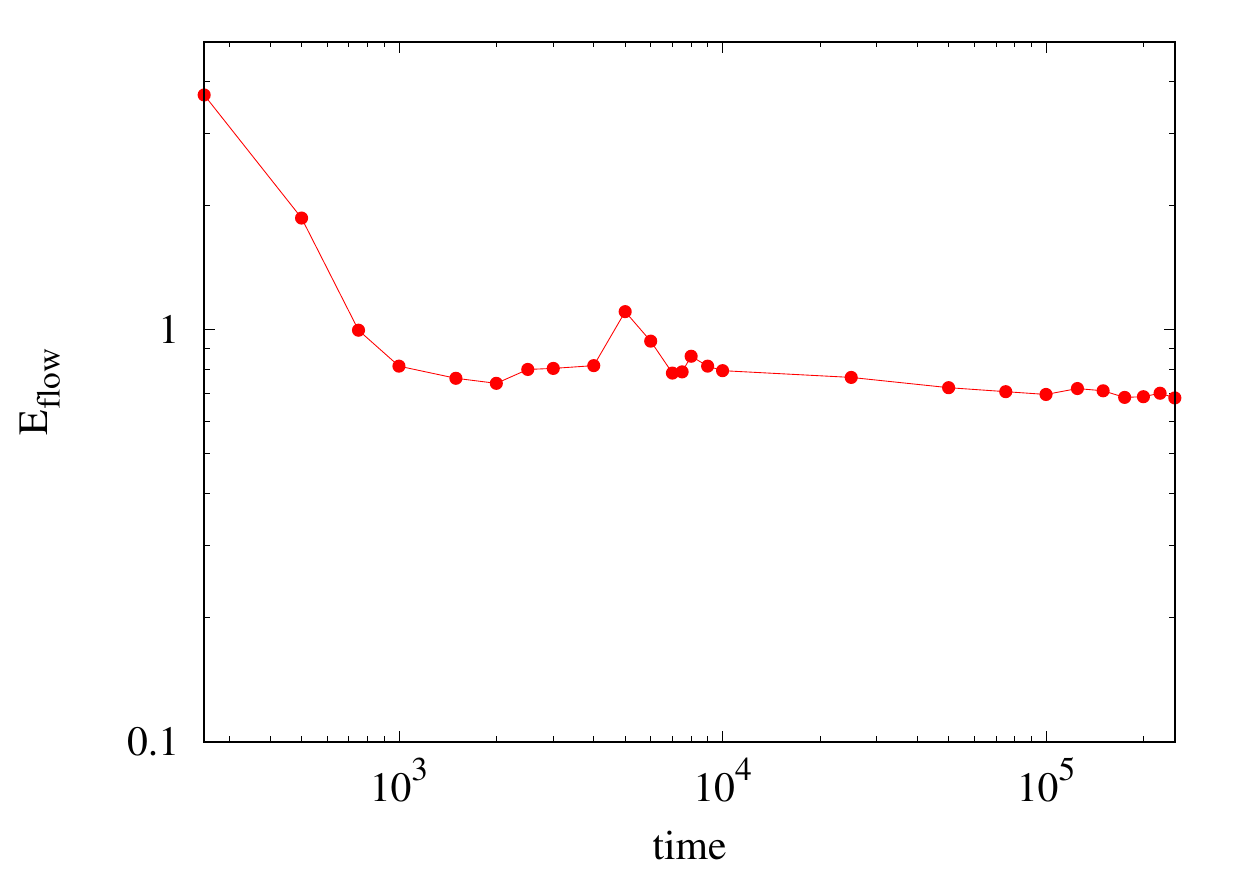}
 \caption{(Color online) The variation of $E_{{\rm flow}}$ [see Eq.~(\ref{eq:flow})] with time. The data are for  ambient number density $\rho_0=0.15$. }
 \label{fig:e_flow}
\end{figure}

The total energy of the system can be divided into two parts:  one is from the mean flow velocity and the other is from the fluctuations about the mean flow velocity. The energy associated with the mean flow, $E_{{\rm flow}}$, is defined as
\be
E_{{\rm flow}}= \frac{1}{2} \int \rho_r \langle v_r \rangle^2 2 \pi r  dr,
\label{eq:flow}
\ee
where $\rho_r$ is the mean density and $\langle v_r \rangle$ is the mean radial velocity. Figure~\ref{fig:e_flow} shows the temporal variation of $E_{{\rm flow}}$. It can be seen that $E_{{\rm flow}}$ oscillates and reaches a steady value. The crossover time is similar to the crossover time observed for the  power-law growth of the number of moving particles (see Fig.~\ref{fig:power-law-benchmark}). The fact that $E_{{\rm flow}}$ reaches a steady time independent value shows that ignoring the viscosity term in the Navier Stokes equation in the TvNS theory is a reasonable approximation.

\section{\label{conclusion} Conclusion and discussion}

The main aim of this paper was to resolve the contradiction between the conclusions  of earlier simulations  in two~\cite{mbarbier2016physfluids} and three dimensions~\cite{jpjoy2018arxiv} of shock propagation in hard sphere gases that are  initially  at rest. It had been found that in two dimensions 
the simulation data are consistent with the predictions of hydrodynamics for low to medium densities except for a small difference in the discontinuities at the shock front, and a slight discrepancy near the shock center~\cite{mbarbier2016physfluids}. Contrary to this, in three dimensions the simulation data was inconsistent with the predictions of hydrodynamcis at most spatial locations, ranging from the shock center to the shock front~\cite{jpjoy2018arxiv}. In this paper, we revisit the problem in two dimensions by performing large scale event driven simulations. Conclusions from our simulations are inconsistent with those from earlier simulations, but agrees qualitatively with the results of  simulations in three dimensions. In particular, we find that the simulation data in two dimensions are not consistent with  the TvNS solution.  In particular, the exponents characterising the power law behavior of both temperature and density near the shock center are different in theory and simulations.

We also checked the different assumptions implicit in the TvNS theory within simulations of the hard sphere gas in two dimensions. A key assumption is that of local equilibrium which has the consequence that the local pressure, density and temperature are related through an EOS. We find that the simulation data for all distances between the shock front and shock centre are consistent with the EOS of the hard sphere gas, except for a small deviation near the shock front  [see Fig.~\ref{fig:eqn_of_state}]. Local equilibrium also implies that the velocity fluctuations are gaussian. However, we find that distribution of the fluctuations of the radial velocity is non-gaussian, in particular it has non-zero skewness and skewed towards positive fluctuations. Whether this lack of local equilibrium is the cause of the discrepancy between simulation and theory can be determined by studying a system where the local velocities are reassigned at a constant rate consistent with a Maxwell-Boltzmann distribution with width determined by the local temperature. This is a promising area for future study. It is also quite possible that including the effects of heat conduction is important. While heat conduction is irrelevant in the scaling limit, it imposes the boundary condition that the heat flux is zero at the shock centre. This boundary condition results in zero temperature gradient, as seen in the simulations. Whether including the effects of heat conduction in the hydrodynamic equations will be able to reproduce the simulation results requires a detailed numerical solution of the hydrodynamic equations, which is beyond the scope of this paper.

\section*{Acknowledgments}
The simulations were carried out on the supercomputer Nandadevi at The Institute of Mathematical Sciences.

\section*{References}
\bibliographystyle{iopart-num}

\begin{thebibliography}{10}
\expandafter\ifx\csname url\endcsname\relax
  \def\url#1{{\tt #1}}\fi
\expandafter\ifx\csname urlprefix\endcsname\relax\def\urlprefix{URL }\fi
\providecommand{\eprint}[2][]{\url{#2}}

\bibitem{Whithambook}
Whitham G 1974 {\em Linear and Nonlinear Waves\/} (New York: Wiley)

\bibitem{barenblatbook}
Barenblatt G 1987 {\em Scaling, Self-similarity, and Intermediate Asymptotics:
  Dimensional Analysis and Intermediate Asymptotics\/} (Cambridge: Cambridge
  University Press)

\bibitem{gtaylor1950}
Taylor G 1950 {\em Proc. R. Soc. Lond. A\/} {\bf 201} 159

\bibitem{gtaylor1950_2}
Taylor G 1950 {\em Proc. R. Soc. Lond. A\/} {\bf 201} 175--186

\bibitem{jvneumann1963cw}
von Neumann J 1963 {\em Collected Works\/} (Oxford: Pergamon Press) p 219

\bibitem{lsedov_book}
Sedov L 1993 {\em Similarity and Dimensional Methods in Mechanics\/} 10th ed
  (Florida: CRC Press)

\bibitem{sedov1946}
Sedov L 1946 {\em J. Appl. Math. Mech.\/} {\bf 10} 241

\bibitem{lwoltjer1972}
Woltjer L 1972 {\em Ann. Rev. Astron. Astrophys.\/} {\bf 10} 129--158

\bibitem{sfgull1973}
Gull S 1973 {\em Mon. Not. R. Astr. Soc.\/} {\bf 161} 47--69

\bibitem{dfcioffi1988}
Cioffi D~F, Mckee C~F and Bertschinger E 1988 {\em The Astrophysical Journal\/}
  {\bf 334} 252--265

\bibitem{jpostriker1988rmp}
Ostriker J~P and McKee C~F 1988 {\em Rev. Mod. Phys.\/} {\bf 60}(1) 1--68

\bibitem{ybzeldovich_book}
Zel'dovich Y~B and Raizer Y~P 2002 {\em Physics of Shock Waves and High
  Temperature Hydrodynamic Phenomena\/} (New York: Dover Publications, Inc.)

\bibitem{mjedwards2001prl}
Edwards M~J, MacKinnon A~J, Zweiback J, Shigemori K, Ryutov D, Rubenchik A~M,
  Keilty K~A, Liang E, Remington B~A and Ditmire T 2001 {\em Phys. Rev.
  Lett.\/} {\bf 87}(8) 085004

\bibitem{adedensphysplasma2004}
Edens A, Ditmire T, Hansen J, Edwards M, Adams R, Rambo P, Ruggles L, Smith I
  and Porter J 2004 {\em Phys. Plasmas\/} {\bf 11}(11) 4968--4972

\bibitem{asmoorephysplasma2005}
Moore A~S, Symes D~R and Smith R~A 2005 {\em Phys. Plasmas\/} {\bf 12}(05)
  052707--1--052707--7

\bibitem{dokuchaev2002aanda}
Dokuchaev V~I 2002 {\em Astronomy and Astrophysics\/} {\bf 395}(3) 1023--1029

\bibitem{saegfalle1975astronandastrophys}
Falle S 1975 {\em Aston. and Astrophys.\/} {\bf 43} 323--336

\bibitem{afghoniem1982jfm}
Ghoniem A, Kamel M, Berger S and Oppenheim A 1982 {\em J. Fluid Mech\/} {\bf
  117} 473--491

\bibitem{amraouf1991fluiddynres}
Abdel-Raouf A and Gretler W 1991 {\em Fluid Dyn. Res.\/} {\bf 8} 273--285

\bibitem{hsteiner1994physfluids}
Steiner H and Gretler W 1994 {\em Phys. Fluids\/} {\bf 6} 2154

\bibitem{neumann1950japplphys}
VonNeumann J and Richtmyer R 1950 {\em Journal of Applied Physics\/} {\bf 21}
  232

\bibitem{rlatter1955japplphys}
Latter R 1955 {\em Journal of Applied Physics\/} {\bf 26} 954

\bibitem{hlbrode1955japplphys}
Brode H~L 1955 {\em Journal of Applied Physics\/} {\bf 26} 766

\bibitem{mnplooster1970physfluids}
Plooster M~N 1970 {\em The Phys. Fluids\/} {\bf 13} 2665

\bibitem{amwalsh2003prl}
Walsh A~M, Holloway K~E, Habdas P and de~Bruyn J~R 2003 {\em Phys. Rev.
  Lett.\/} {\bf 91}(10) 104301

\bibitem{ptmetzger2009aip}
Metzger P~T, Latta R~C, Schuler J~M and Immer C~D 2009 {\em AIP Conf. Proc\/}
  {\bf 1145} 767

\bibitem{ygrasselli2001granmatt}
Grasselli Y and Herrmann H~J 2001 {\em Gran Matt\/} {\bf 3} 201--204

\bibitem{jfboudet2009prl}
Boudet J~F, Cassagne J and Kellay H 2009 {\em Phys. Rev. Lett.\/} {\bf 103}(22)
  224501

\bibitem{zjabeen2010epl}
Jabeen Z, Rajesh R and Ray P 2010 {\em Eur. Phys. Lett.\/} {\bf 89} 34001

\bibitem{snpathak2012pre}
Pathak S~N, Jabeen Z, Ray P and Rajesh R 2012 {\em Phys. Rev. E\/} {\bf 85}(6)
  061301

\bibitem{xcheng2008natphy}
Cheng X, Xu L, Patterson A, Jaeger H~M and Nagel S~R 2008 {\em Nat Phys\/} {\bf
  4} 234

\bibitem{bsandnes2007prl}
Sandnes B, Knudsen H~A, M\aa{}l\o{}y K~J and Flekk\o{}y E~G 2007 {\em Phys.
  Rev. Lett.\/} {\bf 99}(3) 038001
  \urlprefix\url{http://link.aps.org/doi/10.1103/PhysRevLett.99.038001}

\bibitem{sfpinto2007prl}
Pinto S~F, Couto M~S, Atman A~P~F, Alves S~G, Bernardes A~T, de~Resende H~F~V
  and Souza E~C 2007 {\em Phys. Rev. Lett.\/} {\bf 99}(6) 068001

\bibitem{ojohnsen2006pre}
Johnsen O, Toussaint R, M\aa{}l\o{}y K~J and Flekk\o{}y E~G 2006 {\em Phys.
  Rev. E\/} {\bf 74}(1) 011301

\bibitem{hhuang2012prl}
Huang H, Zhang F and Callahan P 2012 {\em Phys. Rev. Lett.\/} {\bf 108}(25)
  258001

\bibitem{jpjoy2017pre}
Joy J~P, Pathak S~N, Dibyendu D and Rajesh R 2017 {\em Phys. Rev. E\/} {\bf
  96}(3) 032908

\bibitem{mbarbier2015prl}
Barbier M, Villamaina D and Trizac E 2015 {\em Phys. Rev. Lett.\/} {\bf
  115}(21) 214301

\bibitem{mbarbier2016physfluids}
Barbier M, Villamaina D and Trizac E 2016 {\em Phys. Fluids\/} {\bf 28} 083302

\bibitem{tantal2008pre}
Antal T, Krapivsky P~L and Redner S 2008 {\em Phys. Rev. E\/} {\bf 78}(3)
  030301

\bibitem{jpjoy2018arxiv}
Joy J~P, Pathak S~N and Rajesh R 2018 {\em arXiv:1812.03638
  [cond-mat.stat-mech]\/}

\bibitem{dcrapaportbook}
Rapaport D~C 2004 {\em The art of molecular dynamics simulations\/} (Cambridge:
  Cambridge University Press)

\bibitem{masaharumdsimulation2016}
Isobe M 2016 {\em Molecular Simulation\/} {\bf 42}(16) 1317--1329

\bibitem{Landaubook}
Landau L and Lifshitz E 1987 {\em Course of Theoretical Physics- Fluid
  Mechanics\/} (Oxford: Butterwörth-Heinemann)

\bibitem{McCoybook}
McCoy B~M 2009 {\em Advanced Statistical Mechanics\/} (Oxford: Oxford Science
  Publications)

\end{thebibliography}

\end{document}